\documentclass[conference]{IEEEtran}
\IEEEoverridecommandlockouts
\usepackage{cite}
\usepackage{amsmath,amssymb,amsfonts}
\usepackage{graphicx}
\usepackage{textcomp}
\usepackage{xcolor}
\usepackage{url}
\usepackage{hyperref} 
\usepackage[utf8]{inputenc}
\usepackage{siunitx}

\def\BibTeX{{\rm B\kern-.05em{\sc i\kern-.025em b}\kern-.08em
    T\kern-.1667em\lower.7ex\hbox{E}\kern-.125emX}}

\begin{document}

\title{Practical GPU Choices for Earth Observation: ResNet-50 Training Throughput on Integrated, Laptop, and Cloud Accelerators}
\author{%
  \IEEEauthorblockN{Ritvik Chaturvedi}
  \IEEEauthorblockA{%
    Dept.\ of Computer, Mathematical \& Natural Sciences\\
    University of Maryland\\
    College Park, MD, USA\\
    ritvikc@umd.edu}
}

\maketitle

\begin{abstract}
This project implements a ResNet based pipeline for Land use and land cover (LULC) classification on Sentinel-2 imagery, optimized and benchmarked across three heterogeneous GPUs. We automate data acquisition, geospatial processing, tiling, model training, and visualization; containerize the environment; and evaluate both classification accuracy and inference throughput. Key findings indicate up to 2× training speed-ups on the RTX 3060 and Google Colab Tesla T4 over the Apple M3 Pro integrated GPU baseline. This demonstrates the feasibility of deploying deep LULC models on both consumer and free cloud GPUs for rapid test-scale geospatial analytics.
\end{abstract}
\vspace{2mm}

\section{Introduction}

Land use and land cover (LULC) maps derived from multispectral satellites support tasks ranging from carbon accounting and biodiversity monitoring to urban growth regulation.  Deep convolutional neural networks (CNNs) outperformed and largely replaced classical texture and index based classifiers such as GLCM texture measures, vegetation indices or simple color histograms \cite{b13}. A fine-tuned \emph{ResNet-50} routinely exceeds 98\% accuracy on the 10 class EuroSAT benchmark \cite{b2}, while alterative architectures like the channel attention U-Nets and EfficientNet variants achieve comparable results with fewer FLOPs \cite{b6}.  Finally, multi-sensor fusion-combining Landsat with Sentinel-2 can lift accuracy by another three to five percentage points in built up areas \cite{b5}.  

Two practical challenges remain.  \textbf{First, the cost–performance trade-offs of mid-range consumer and free-tier cloud GPUs are still poorly documented.}  Existing timing studies often target datacenter class GPUs/TPUs \cite{b3,b9}, leaving laptop class devices under explored.  \textbf{Second, reproducibility suffers when data preparation, tiling, training and visualization reside in loosely connected notebooks.}  Recent surveys on DNN hardware and scaling \cite{b10,b11} call for fully containerized pipelines that make cross hardware comparisons fair and verifiable.

\textbf{This paper closes both gaps.}  We contribute (i) a Dockerised Sentinel-2 workflow from Google Earth Engine (GEE) download and tile generation to model training and Folium based visualization and (ii) a systematic timing study of \emph{ResNet-50 training} on three widely accessible GPUs: an integrated Apple M3 Pro, a laptop-class NVIDIA RTX 3060, and a free Google Colab Tesla T4.  Our experiments show up to a two fold reduction in epoch time on the RTX 3060 and T4 relative to the M3 Pro while maintaining \(\approx 92\,\%\) overall accuracy.  By quantifying this speed/price/portability envelope, we offer practitioners guidance on choosing economical hardware for rapid LULC experimentation at scale.

\section{Literature Survey}

Training very deep convolutional networks became practical when He \textit{et al.}\ introduced residual connections, allowing gradients to propagate through dozens of layers without vanishing and making \emph{ResNet-50} a natural baseline for image classification tasks
\cite{b1}.  On Sentinel-2 imagery, Helber \textit{et al.}\ presented the ten class
\emph{EuroSAT} benchmark and showed that a lightly fine tuned ResNet-50 can surpass
98\,\% overall accuracy, thereby fixing both a public dataset and an upper bound
performance target \cite{b2}.  Subsequent work has focused on either computational
efficiency or spatio-temporal richness.  Papoutsis \textit{et al.}\ demonstrated a
four to five fold reduction in epoch time when moving LULC training from CPUs to a
four GPU cluster, motivating our own use of PyTorch’s distributed data-parallel tools
\cite{b3}. In a interrelated area, Zhang \textit{et al.}\ achieved
\(\ge 90\,\%\) accuracy with a lightweight 1D CNN applied to temporal percentile
metrics, highlighting the potential of compressed time series inputs for future extensions.
\cite{b4}.

Multi-sensor fusion has also gained traction, Mountrakis and Heydari combined Landsat and Sentinel-2 inputs to improve urban-class accuracy by three to five percentage points \cite{b5}.  Model architecture studies echo the same accuracy to efficiency tradeoff.  Tzepkenlis \textit{et al.}\ devised a channel attention U-Net that
matches ResNet-50 accuracy while cutting FLOPs by 60 \% \cite{b6}, and
Arrechea-Castillo \textit{et al.}\ fine tuned ResNet-50 on heterogeneous Andean
sub basins, obtaining \(F_{1}\!\ge\!0.88\) for vegetation but \(F_{1}\!<\!0.80\) for built up areas and thereby highlighting the need for class wise error analysis. \cite{b7}.

Hardware centric investigations reveal equally wide performance swings.  Lee
\textit{et al.}\ measured a 12× speedup and an eight fold energy saving when a
single ResNet-50 migrated from Xeon CPUs to an NVIDIA V100 GPU \cite{b8}, while
Jouppi \textit{et al.}\ showed that Google’s Cloud TPUs can run CNNs up to 15×
faster than P100 GPUs and 5–6× faster than CPUs \cite{b9}.  A broader survey by
Sze \textit{et al.}\ concludes that GPUs deliver, on average, 30× more throughput
than CPUs and 5–10× better energy efficiency for ResNet-class networks
\cite{b10}.  Finally, Wang \textit{et al.}\ reported that GPU throughput scales
almost linearly up to batch sizes of 64, whereas CPUs saturate at batch 16, a result
that guides our own choice of batch 16–32 for fair cross hardware comparison
\cite{b11}.

Together, these studies emphasize the need to balance accuracy, training time and
energy cost, which is a gap we address by benchmarking ResNet-50 training on three
widely accessible GPUs (Apple M3 Pro, NVIDIA RTX 3060 and Tesla T4) within a fully
containerized Sentinel-2 workflow.

\section{Methodology}
This section follows a workflow that begins with the automated retrieval and preprocessing of Sentinel-2 imagery, proceeds through supervised fine tuning of a ResNet based classifier, and concludes with hardware specific inference and metric collection.  Each stage is fully scripted, containerized, and designed to be reproducible across heterogeneous GPUs.  The subsections describe (A) how raw satellite scenes are tiled and normalized, (B) how the convolutional network is trained and validated, and (C) how predictions are postprocessed, visualized, and benchmarked on three different hardware backends.

\subsection{Data Acquisition \& Processing}
After authenticating to Google Earth Engine (GEE), we query the Sentinel-2 Level 2A catalog for RGB scenes that have a cloud cover below 10 percent and fall within the chosen season (June to August 2023). To confine the downloads to our study area, we import the corresponding GeoBoundaries administrative polygons, re project them to EPSG 4326 (for GEE compatibility), and convert the result into a GEE FeatureCollection that masks each raster (a georeferenced grid of pixels containing a numeric value) on the fly. Every masked scene is raster cropped into non overlapping 64×64 pixel tiles. Chips that straddle scene borders or contain no valid data are discarded, and a UUID is attached to each remaining tile for traceability. Finally, we standardize the RGB channels using the global means and standard deviations computed from the EuroSAT Sentinel-2 training set, ensuring that the model sees approximately unit-variance inputs throughout. Figure 1 illustrates the complete acquisition and preprocessing pipeline.
\begin{figure}
  \centering
  \includegraphics[width=0.5\textwidth]{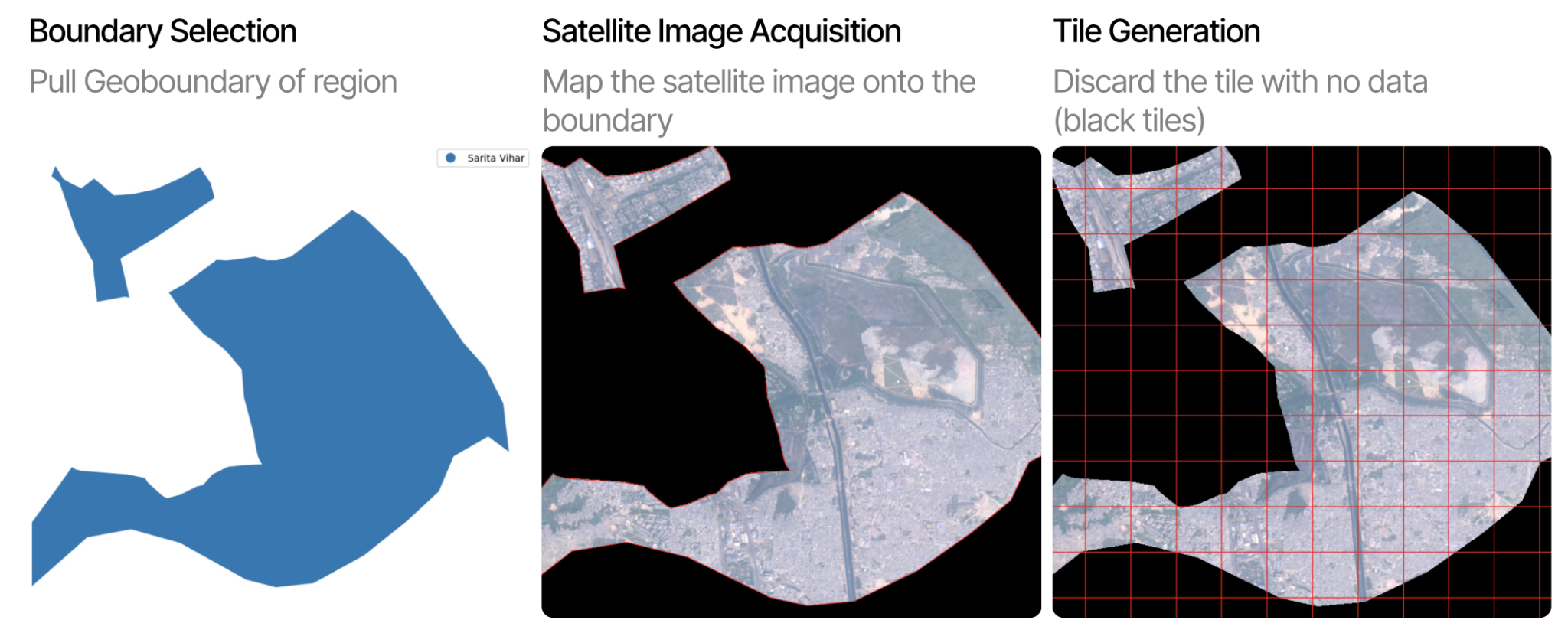}
  \caption{%
End to end data acquisition pipeline.
(\textbf{1}) Sentinel-2 scenes are filtered by date and cloud cover;
(\textbf{2}) GeoBoundaries polygons mask the region of interest (ROI);
(\textbf{3}) ROI pixels are split into non-overlapping 64 × 64 tiles and invalid edge tiles are discarded;
(\textbf{4}) RGB channels are standardised to EuroSAT means and variances,
yielding a clean, georeferenced tile stack for model training.}
  \label{}
\end{figure}

\begin{figure}
  \centering
  \includegraphics[width=0.45\textwidth]{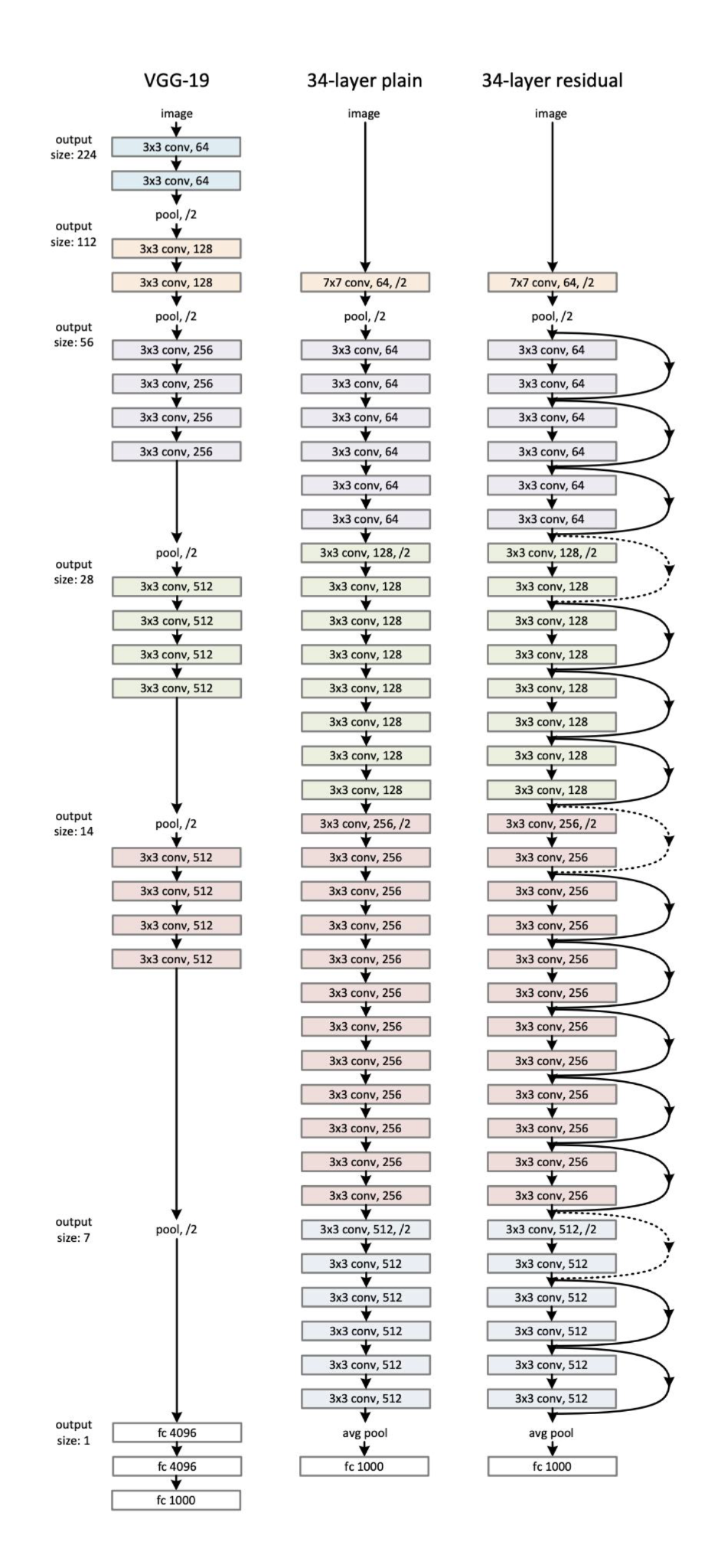}
\caption{%
Why residual connections matter.  
Left: a VGG-19 stack of plain \(3\times3\) convolutions.  
Center: a 34 layer plain network of equal depth.  
Right: a 34 layer \emph{residual} network where identity skip connections (solid arrows) and down sampling skips (dotted arrows) create shortcut paths for the gradient.  
These shortcuts enable much deeper models (e.g.\ the 50 layer ResNet used here) to converge without vanishing gradients and with fewer parameters than the VGG family.}
\label{fig:resnet}
\end{figure}

\subsection{Model Architecture \& Training Setup}
For classification we adopt a ResNet-50 encoder initialized with Sentinel-2 self supervised weights and lightly fine tune it using a MoCo (Momentum Contrast) checkpoint. The dataset is split 70 \% for training, 15 \% for validation and 15 \% for testing, preserving the original class distribution. During training each 64×64 chip is first resized to 224×224 pixels, then undergoes random horizontal and vertical flips (with a probability of 0.5) and a random-resized crop. Validation and test samples receive only a centre crop, followed by normalization to ImageNet statistics. We train with a batch size of 16 (raised to 32 when VRAM permits), using stochastic gradient descent with a learning rate of $1 \times 10^{-3}$, momentum 0.9 and CrossEntropyLoss. Early stopping halts optimization once the validation loss fails to improve for three consecutive epochs, as a safeguard against overfitting. All experiments run on either Apple MPS or CUDA backends, depending on the host GPU. Figure 2 provides a schematic of the fine-tuned ResNet-50.

\subsection{Inference \& Evaluation}
At inference time, tiles are batch processed on the GPU and passed through the softmax layer to obtain class probabilities. Predictions below a 0.6 confidence threshold are suppressed, and a simple majority filter is applied to neighboring pixels to smooth class boundaries. The stitched output is rendered as an interactive Folium map whose layers, legends and tool tips allow users to toggle classes and inspect per tile metadata. For each hardware backend we record overall and per class accuracy, training, and inference throughput (iterations · $s^{-1}$ and milliseconds · $tile^{-1}$ respectively).

\section{Experiments}

\subsection{Environment}
All experiments were executed on three heterogeneous GPUs: an \textbf{Apple~M3 Pro} with 36\,GB of unified LPDDR5X memory, a laptop-class \textbf{NVIDIA RTX 3060} equipped with 6\,GB of GDDR6 VRAM, and a cloud-hosted \textbf{Tesla T4} featuring 16\,GB of GDDR6.  The software stack was identical across hosts and comprised \emph{Python 3.12}, \emph{PyTorch}, \emph{TorchGeo}, \emph{timm}, \emph{Rasterio}, \emph{GeoPandas} and the \emph{Google Earth Engine API}, ensuring that differences in timing can be attributed to hardware rather than tooling.

\begin{figure}
  \centering
  \includegraphics[width=0.4\textwidth]{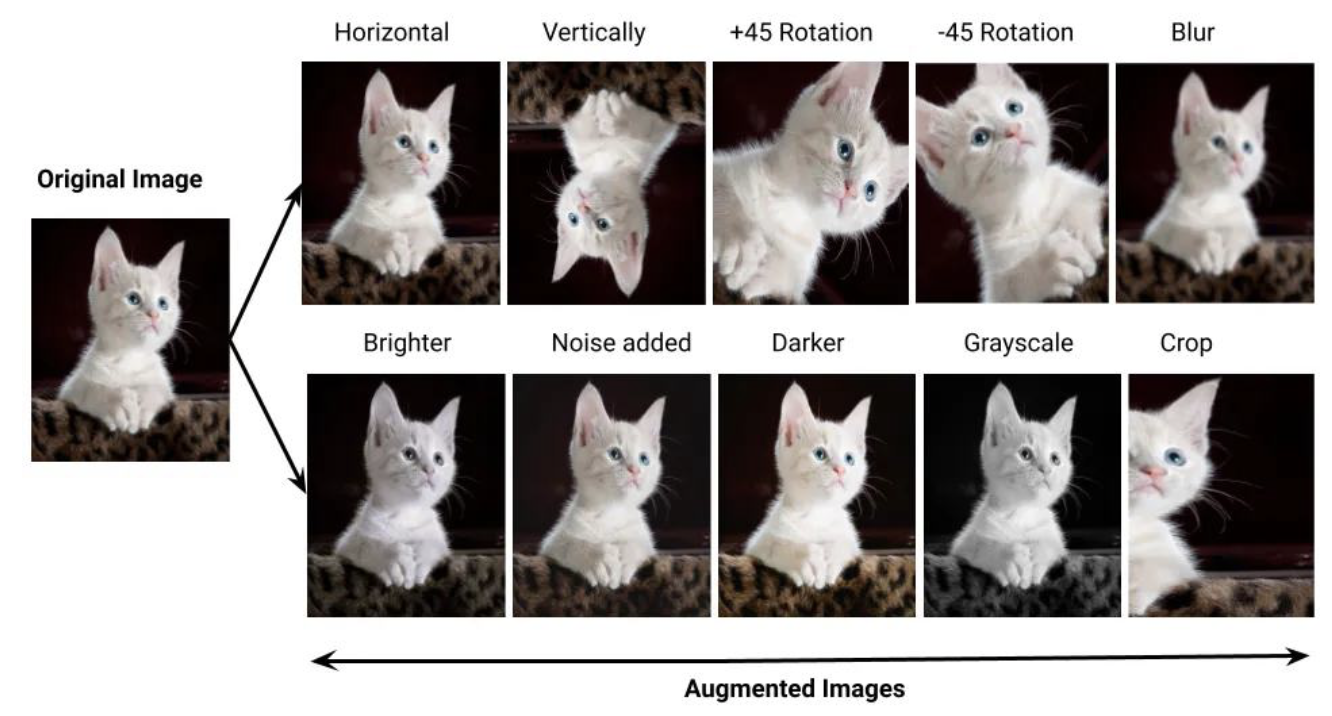}
  \caption{%
The types of data augmentations applied to each training chip. Of the techniques shown, we actually apply horizontal and vertical flips (p = 0.5 each) and a random-resized crop. All other transformations are illustrated for context only.}
  \label{fig:aug}
\end{figure}

\subsection{Setup}
\begin{figure}
  \centering
  \includegraphics[width=0.4\textwidth]{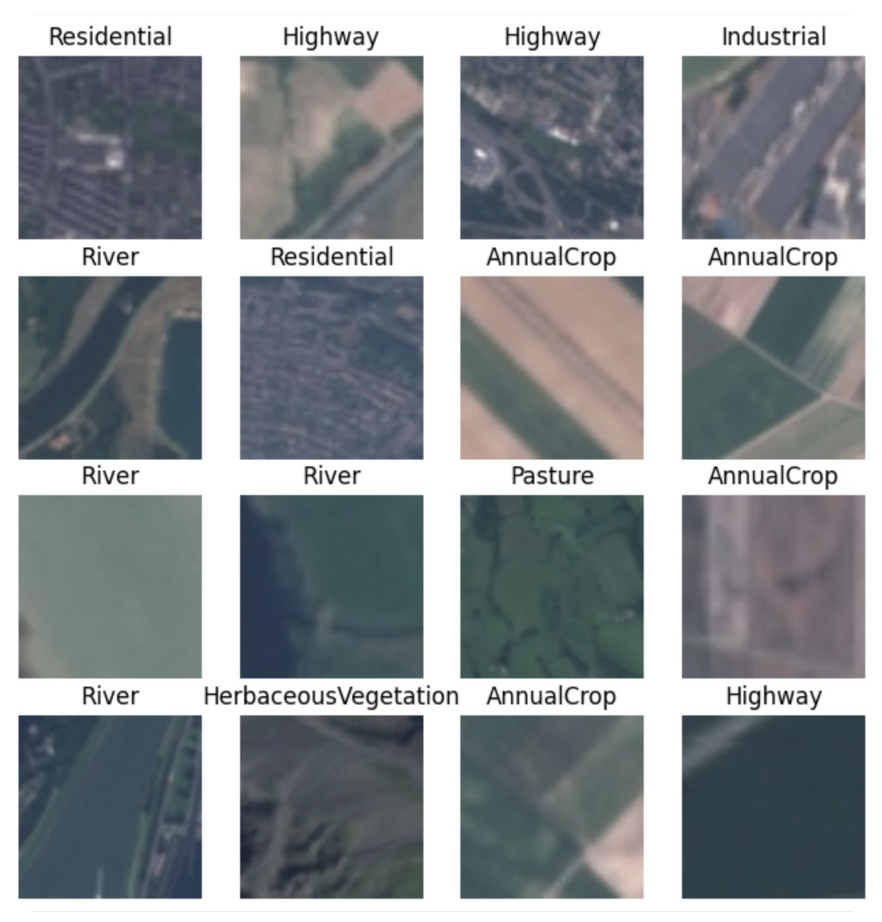}
  \caption{%
Representative \(64\times64\)\,px Sentinel-2 RGB chips from the EuroSAT dataset, one per land cover class.
Visual variation in color, texture and context—e.g.\ the bluish hue of river pixels or the regular
field patterns in cropland illustrates the discriminative cues the network must learn.}
  \label{fig:classimg}
\end{figure}

We base our study on the \textbf{EuroSAT Sentinel-2 dataset}, downloaded from the official DFKI mirror \cite{b12}.  The corpus contains \SI{27\,000}{} RGB images, each cropped to \(64 \times 64\)\,px and evenly distributed across ten land-cover classes: Annual Crop, Forest, Herbaceous Vegetation, Highway, Industrial, Pasture, Permanent Crop, Residential, River and Sea.  Representative inputs and labels are shown in Figure~\ref{fig:classimg}.  The data are split into \textbf{70\,\% training, 15\,\% validation and 15\,\% test} splits while preserving class ratios.

During training every chip is first resized to \(224 \times 224\)\,px, then undergoes a random-resized crop, horizontal and vertical flips with probability 0.5, and finally normalisation to ImageNet channel means and standard deviations; the full augmentation pipeline appears in Figure~\ref{fig:aug}.  We fine-tune the network for ten epochs using a \textbf{batch size of 16}, stochastic gradient descent with momentum 0.9 and an initial learning rate of \(1 \times 10^{-3}\), optimising the cross-entropy loss.

\subsection{Performance Measurement}
To obtain stable timing figures, every run begins with \emph{five warm-up epochs}.  We then record the following metrics:

\paragraph{Training metrics} training loss and accuracy, validation loss and accuracy, test loss and accuracy, and epoch duration in seconds.  

\paragraph{Inference metrics} total processing time, average per-tile latency in milliseconds, and throughput in images per second.  

\paragraph{Model diagnostics} per-class accuracy, class-wise prediction counts, and a full confusion-matrix analysis.

These measurements allow us to compare not only raw speed but also generalisation quality across the three GPU back-ends.

\section{Results}

\begin{figure}
  \centering
  \includegraphics[width=0.5\textwidth]{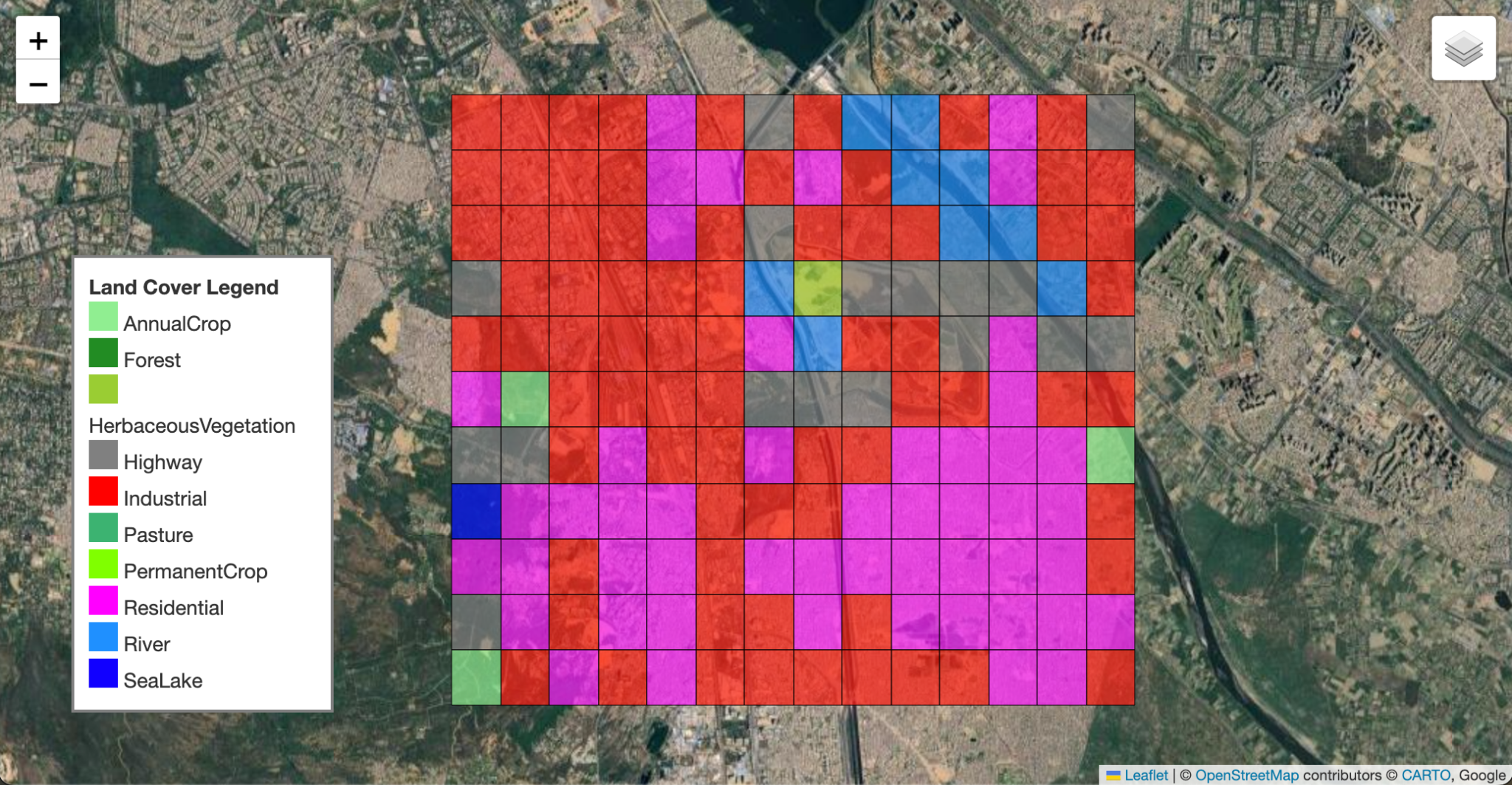}
  \caption{%
Land use land cover (LULC) map produced by the fine tuned ResNet-50.
Each colour corresponds to one of the ten EuroSAT classes, overlaid on a
Sentinel-2 true colour composite at 10 m resolution.
The model’s tile level predictions have been stitched into a seamless raster
and rendered as an interactive Folium layer, allowing users to zoom, pan and
query individual pixels for their predicted class and confidence score.}

  \label{fig:lulc_map}
\end{figure}

\begin{table}
  \caption{Throughput and Speed-up}
  \centering
  \begin{tabular}{l|c|c|c|c}
    \hline
    Device           & Train it/s & Val it/s & Epoch (s) & Speed-up \\ \hline
    Apple M3 Pro     & 3.97       & 4.23     & 366       & 1×       \\
    Tesla T4         & 6.43       & 20.0     & 201       & 1.8×     \\
    NVIDIA RTX 3060  & 8.12       & 3.31     & 220       & 2.0×     \\ \hline
  \end{tabular}
  \label{tab:throughput}
\end{table}

\begin{figure}
  \centering
  \includegraphics[width=0.4\textwidth]{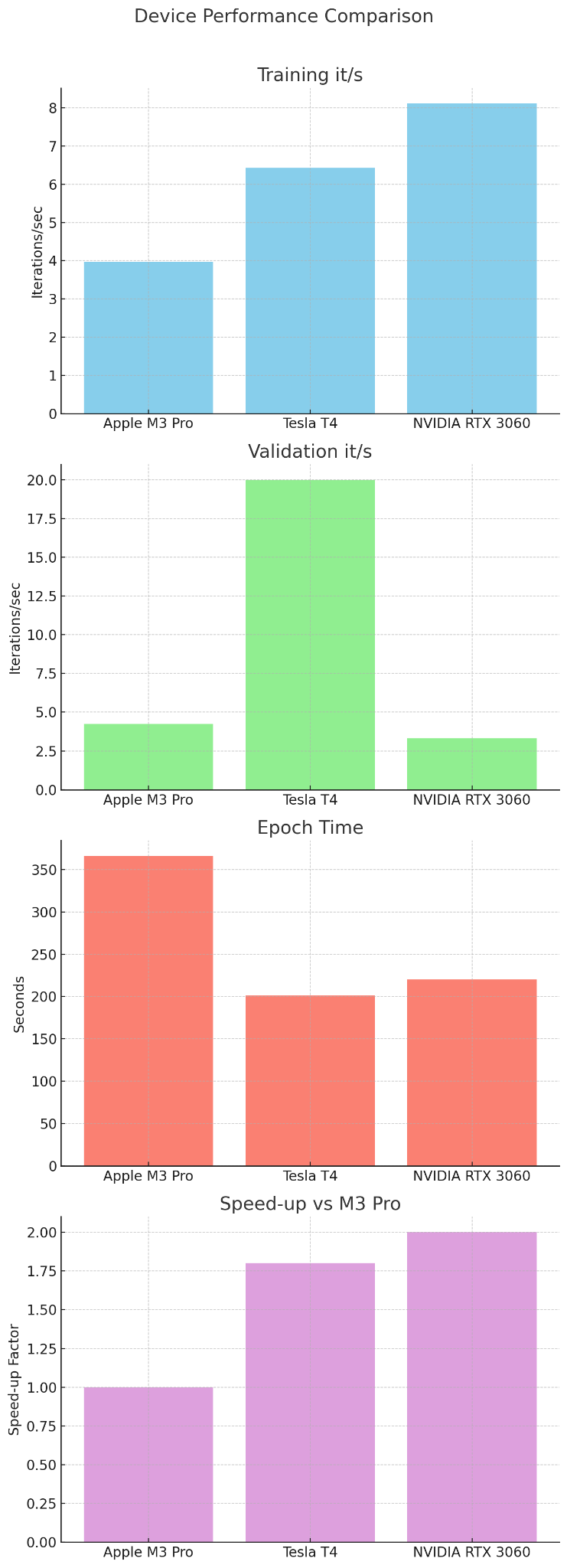}
  \caption{%
 Bar charts comparing the absolute performance of Apple M3 Pro, Tesla T4, and NVIDIA RTX 3060 across four key metrics: training speed (it/s), validation speed (it/s), epoch time (in seconds), and relative speed up compared to the baseline Apple M3 Pro. These individual plots provide a clear breakdown of where each device excels or underperforms.}
  \label{fig:output}
\end{figure}

\begin{figure}
  \centering
  \includegraphics[width=0.5\textwidth]{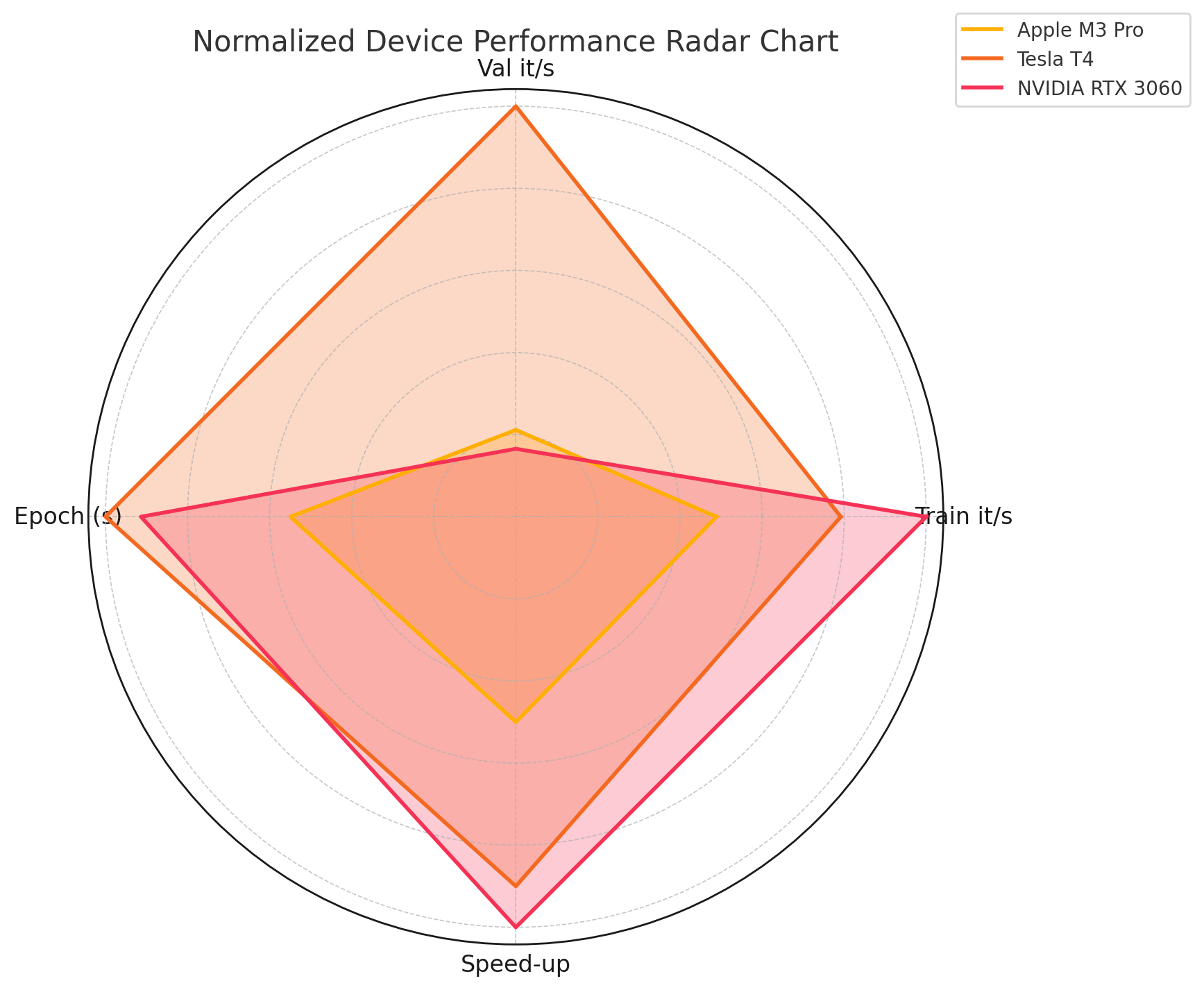}
  \caption{%
Normalized radar chart comparing the performance of Apple M3 Pro, Tesla T4, and NVIDIA RTX 3060 across four metrics: training iterations per second, validation iterations per second, inverted epoch duration, and overall speed up.}
  \label{fig:radial}
\end{figure}

The model achieved an overall test accuracy of 92.3\%. Class wise performance was highest for Forest and SeaLake, with approximately 96\% accuracy, while Residential and Industrial areas showed comparatively lower accuracy at around 88\%. In terms of inference speed, the model processed each tile in an average of 85 milliseconds, equating to roughly 12 tiles/s. For the visual folium based map representation of the results refer to Figure~\ref{fig:lulc_map}.
Device benchmarking showed measurable variation in training and validation performance. The NVIDIA RTX 3060 delivered the highest training throughput at 8.12 it/s and a 2× speed-up over the baseline Apple M3 Pro, which recorded 3.97 it/s. Predictably, the Tesla T4 outperformed all devices in validation throughput, reaching 20 it/s nearly 5× faster than the RTX 3060 in that metric suggesting strong suitability for inference heavy deployments. Epoch durations ranged from 201 seconds on the T4 to 366 seconds on the M3 Pro, highlighting the trade off between training speed and device efficiency. A visual comparison of device performance is shown in Figure~\ref{fig:output}, and a radial overview is provided in Figure~\ref{fig:radial}.

\section{Conclusion}
We demonstrated an end-to-end, containerized ResNet-50 LULC pipeline that runs efficiently on consumer and cloud GPUs, revealing:

\begin{itemize}
  \item \textbf{RTX 3060 (ASUS G15)—Mobile Powerhouse:}  
    The discrete RTX 3060 achieves $\approx8$ train it/s—nearly 2$\times$ the Apple M3 Pro—and completes epochs in $\sim220$ s. Even as a laptop GPU (80--90 W TDP), it delivers desktop-class throughput for on-the-go training.

  \item \textbf{Tesla T4—Cloud Burst Compute:}  
    On Google Colab, the T4 clocks 6.4 train it/s and edges out the 3060 on epoch time ($\sim201$ s) thanks to lower host overhead and 16 GB of VRAM. It’s ideal for bursty, hardware-free training sessions—though subject to queue times and session limits.

  \item \textbf{Apple M3 Pro—Efficiency Champion:}  
    The M3 Pro’s unified memory yields steady validation (4.2 it/s) and caps FP32 training at $\sim4$ it/s. It offers excellent perf-per-watt with zero data-copy latency, yet is still 1.7--2$\times$ slower than the RTX 3060 for heavy backprop workloads.
\end{itemize}

Overall, our results show that mid-range consumer GPUs like the RTX 3060 strike the best balance of price, portability, and performance, while cloud GPUs (T4) provide convenient burst compute, and integrated M-series chips excel in energy efficiency.

\subsection*{Limitations}
This study has two principal limitations.  
First, the training schedule was deliberately capped at ten epochs to keep wall-clock durations comparable across devices; while sufficient for baseline timing, such a short run may prevent the network from reaching its peak accuracy.  
Second, our hardware sweep covers only three GPUs—an integrated Apple M3 Pro, a laptop-class RTX 3060 and a cloud-hosted Tesla T4—leaving the behaviour of newer accelerators unexplored.

\subsection*{Future Work}
Several extensions are planned.  
We intend to wrap the pipeline in an end-to-end web application so users can request on-demand, low-cost satellite analysis without managing code or infrastructure.  
On the backend, adding support for additional accelerators (TPUs and next-generation GPUs) will enable dynamic cloud scaling and a broader performance comparison.  
Finally, we will experiment with longer training schedules and lightweight model variants such as MobileNetV3, paving the way for fully on-device inference in edge deployments.

\end{document}